\documentclass[aip,jap,amsmath,amssymb,reprint,superscriptaddress]{revtex4-1}
\usepackage{graphicx}
\usepackage{dcolumn}
\usepackage{bm}
\usepackage{gensymb}
 
\begin{document}
\preprint{AIP/123-QED}
\title{Transient reflectance of photoexcited Cd$_3$As$_2$}
\date{\today}
\author{C. P. Weber}\email{cweber@scu.edu}
\affiliation{Dept. of Physics, Santa Clara University, 500 El Camino Real, Santa Clara, CA 95053-0315, USA}
\author{Ernest Arushanov}
\affiliation{Institute of Applied Physics, Academy of Sciences of Moldova, Academiei str. 5, MD 2028 Chisinau, Moldova}
\author{Bryan S. Berggren}
\affiliation{Dept. of Physics, Santa Clara University, 500 El Camino Real, Santa Clara, CA 95053-0315, USA}
\author{Tahereh Hosseini}
\author{Nikolai Kouklin}
\affiliation{Departments of Electrical Engineering and Computer Science, University of Wisconsin-Milwaukee, P.O. Box 413, Milwaukee, Wisconsin 53201}
\author{Alex Nateprov}
\affiliation{Institute of Applied Physics, Academy of Sciences of Moldova, Academiei str. 5, MD 2028 Chisinau, Moldova}

\begin{abstract}
We report ultrafast transient-grating measurements of crystals of the three-dimensional Dirac semimetal cadmium arsenide, Cd$_3$As$_2$, at both room temperature and 80 K. After photoexcitation with 1.5-eV photons, charge-carriers relax by two processes, one of duration 500 fs and the other of duration 3.1 ps. By measuring the complex phase of the change in reflectance, we determine that the faster signal corresponds to a decrease in absorption, and the slower signal to a decrease in the light's phase velocity, at the probe energy. We attribute these signals to electrons' filling of phase space, first near the photon energy and later at lower energy. We attribute their decay to cooling by rapid emission of optical phonons, then slower emission of acoustic phonons. We also present evidence that both the electrons and the lattice are strongly heated.
\\ \\
The following article appeared in \textit{Applied Physics Letters} and may be found at\\ \url{http://scitation.aip.org/content/aip/journal/apl/106/23/10.1063/1.4922528}
\\
(This version of the article differs slightly from the published one.) 
 \\ \\
\textit{Copyright 2015 American Institute of Physics. This article may be downloaded for personal use only. 
Any other use requires prior permission of the author and the American Institute of Physics.}
\end{abstract}
\maketitle

\section{Introduction}

Cadmium arsenide, known for decades as an inverted-gap semiconductor, has recently been shown to be a three-dimensional Dirac semimetal.\protect{\cite{Wang2013,Neupane2014,Liu2014,Borisenko2014}} These materials, with a massless Dirac dispersion throughout the bulk, are the 3D analogs of graphene, and Cd$_3$As$_2$ is foremost among them: stable, high-mobility, and nearly stoichiometric. It displays giant magnetoresistance,\protect{\cite{Liang2015}} hosts topologically nontrivial Fermi-arc states on its surface,\protect{\cite{Yi2014}} and is predicted to serve as a starting point from which to realize a Weyl semimetal, quantum spin Hall insulator, or axion insulator.\protect{\cite{Wan2011,Wang2013}}

Ultrafast spectroscopy, which monitors changes in a sample's optical properties after excitation by a short laser pulse, has in many materials provided a time-resolved probe of basic carrier relaxation processes such as electron-electron and electron-phonon scattering and carrier diffusion. Calculations\protect{\cite{Lundgren2015}} for Dirac and Weyl semimetals predict that photoexcited electrons will, anomalously, cool linearly with time once their elergy drops below that of the lowest optical phonon. Nothing, however, is known of cadmium arsenide's ultrafast properties. Here we use the transient-grating method, which measures both the magnitude and phase of the complex change of reflectance. Our measurements reveal two processes, distinct in lifetime and in phase, by which the sample's reflectance recovers after photoexcitation. Analysis of the signal's phase allows us to identify changes in both the real and the imaginary parts of the index of refraction, $n=n_r+in_i$. The fastest response, with a lifetime of 500 fs, is a reduction in the absorptive part, $n_i$, which we attribute to photoexcited electrons' filling of states near the excitation energy. The longer-lived response is an increase in $n_r$ and arises from the filling of states at much lower energy. These observations reveal a two-stage cooling process, which we suggest may proceed first through optical phonons, then through acoustic. 

\section{Methods}
\subsection{Samples}
We measured two samples of Cd$_3$As$_2$. Sample 1 had well-defined crystal facets and measured a few millimeters in each dimension. It was grown by evaporation of material previously synthesized in Argon flow\protect{\cite{Arushanov1981}} and was annealed at room-temperature for several decades. Such annealing is known to increase electron mobility and to decrease electron concentration.\protect{\cite{Rambo1979}} Indeed, Hall measurements on a sample of the same vintage give electron density $n=6\times10^{17}$ cm$^{-3}$ (roughly independent of temperature), metallic resistivity,\protect{\cite{suppl}} and mobility $\mu=8\times10^4$ cm$^2/\text{V}\,\text{s}$ at 12 K. X-ray powder diffraction gives lattice parameters in agreement with previous reports.\protect{\cite{Arushanov1981}}

Sample 2 was grown in an Argon-purged chamber by CVD in the form of a platelet; the surface was microscopically flat and uniform. The ratio of the main Cd and As peaks seen in energy-dispersive X-ray spectroscopy corresponds to Cd$_3$As$_2$, indicating proper stoichiometry. Though its transport was not unambiguously metallic,\protect{\cite{suppl}} in our experiment samples 1 and 2 behaved identically. This is consistent with the interpretation given below, that our ultrafast signal arises from the dynamics of high-energy electrons.  

\subsection{Transient-grating measurement}
We use the transient-grating method to measure the change, $\Delta r(t)$, in reflectance after photoexcitation. A pair of pump pulses interfere at the sample, exciting electrons and holes in a sinusoidal pattern. The sinusoidal variation in $n$ caused by this excitation is the ``grating.'' Time-delayed probe pulses reflect and diffract off of the grating. 

\begin{figure}
\includegraphics[width=2.7 in, height=0.8 in]{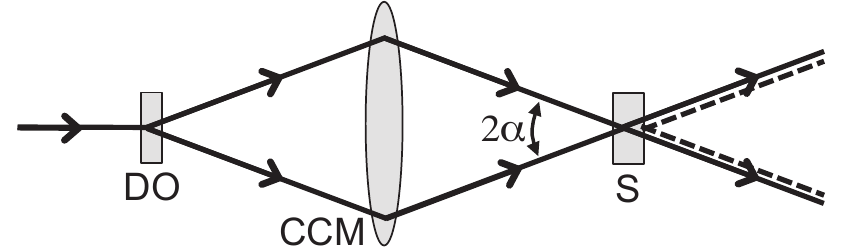}
\caption{Schematic diagram of the transient-grating experiment. For clarity the mirror is shown as a lens, and beams reflected from the sample are shown as transmitted. A probe beam is focused onto a transmissive diffractive optic (DO) that directs most energy into the $\pm1$ orders. A concave mirror (CCM) focuses the two probes onto the sample (S), at an angle $\alpha$ from the normal. Diffracted beams (dashed) scatter through $\pm2\alpha$, so that each diffracted probe is collinear with the opposite reflected probe. Pump beams (not shown) follow the same paths. However, pump beam paths are tipped slightly out of the page, and probe beams slightly into the page. Thus the pumps are not collinear with the probes, nor are the reflected beams collinear with the incident ones. }
\label{tg}
\end{figure}

The experimental geometry is shown in Fig. \ref{tg}. We use a diffractive-optic beamsplitter\protect{\cite{Goodno1998, Maznev1998}} to generate the pair of pump pulses. As these pulses converge on the sample, they make angles $\pm\alpha$ with the surface normal, creating a grating of wavevector $q=(4\pi\sin\alpha)/\lambda$. (Here $\lambda$ is the light's wavelength.) 

Two probe pulses are incident on the sample at the same angles, $\pm\alpha$. The difference in their wavevectors equals $q$, so when each probe diffracts off of the grating, it is scattered to be collinear with the other probe.

This geometry allows for simple heterodyne detection\protect{\cite{Goodno1998, Maznev1998, Gedik2004}} of the diffracted probe: rather than provide a separate ``local oscillator'' beam, the reflected beam from one probe acts as a local oscillator for the diffracted beam from the other probe. If an incident probe has electric field $E_0$, then the reflected and diffracted probe fields are, respectively,
\begin{eqnarray}
E_r & = & |r|e^{i\phi_{r}}E_0+|\Delta r(t)|e^{i\phi_{\Delta r}}E_0,\nonumber\\
E_d & = & |d(t)|e^{i(\phi_{\Delta r}+m\phi_x)}E_0.\label{notation}
\end{eqnarray}
Here $r$ is the complex reflectance, $m$ is the order of diffraction, and $\phi_x$ is a geometric phase due to the grating's spatial location. $\phi_x$ cannot be measured, but it can be changed controllably. Heterodyne detection of $|E_r+E_d|^2$ improves signal, and we suppress noise by modulation of $\phi_x$ and lock-in detection. The transient-grating signal is proportional to\protect{\cite{suppl}}
\begin{equation}
|r||d(t)|\sin(\phi_r-\phi_{\Delta r}-m\phi_x).\label{eq2}
\end{equation}
Each measurement is repeated with the grating shifted by a quarter wavelength, giving the real and imaginary parts of $d(t)$. In the absence of measurable diffusion, as seen here, $d(t)\propto\Delta r(t)$.

The laser pulses have wavelength near 810 nm, duration 120 fs, repetition rate 80 MHz, and are focused to a spot of diameter 114 $\mu$m. The pump pulses have fluence $f$ at the sample of $2.4-9.5$ $\mu$J/cm$^2$; the probe pulses are a factor of 10 weaker. At 810 nm Cd$_3$As$_2$ has index of refraction\protect{\cite{Karnicka1982}} $n=3.3+1.4i$, giving $\phi_r=194\degree$. The absorption length is of order 45 nm and the reflectivity is 35\%, so at our highest fluence each pair of pump pulses photoexcites electrons and holes at a mean density of $n_{\text{ex}}\approx 5.7\times10^{18}$ cm$^{-3}$. Measurements were taken at temperatures $T=295$ K and 80 K, and one at 115 K.

\section{Results}

\begin{figure}
\includegraphics[width=3.375 in, height=1.46 in]{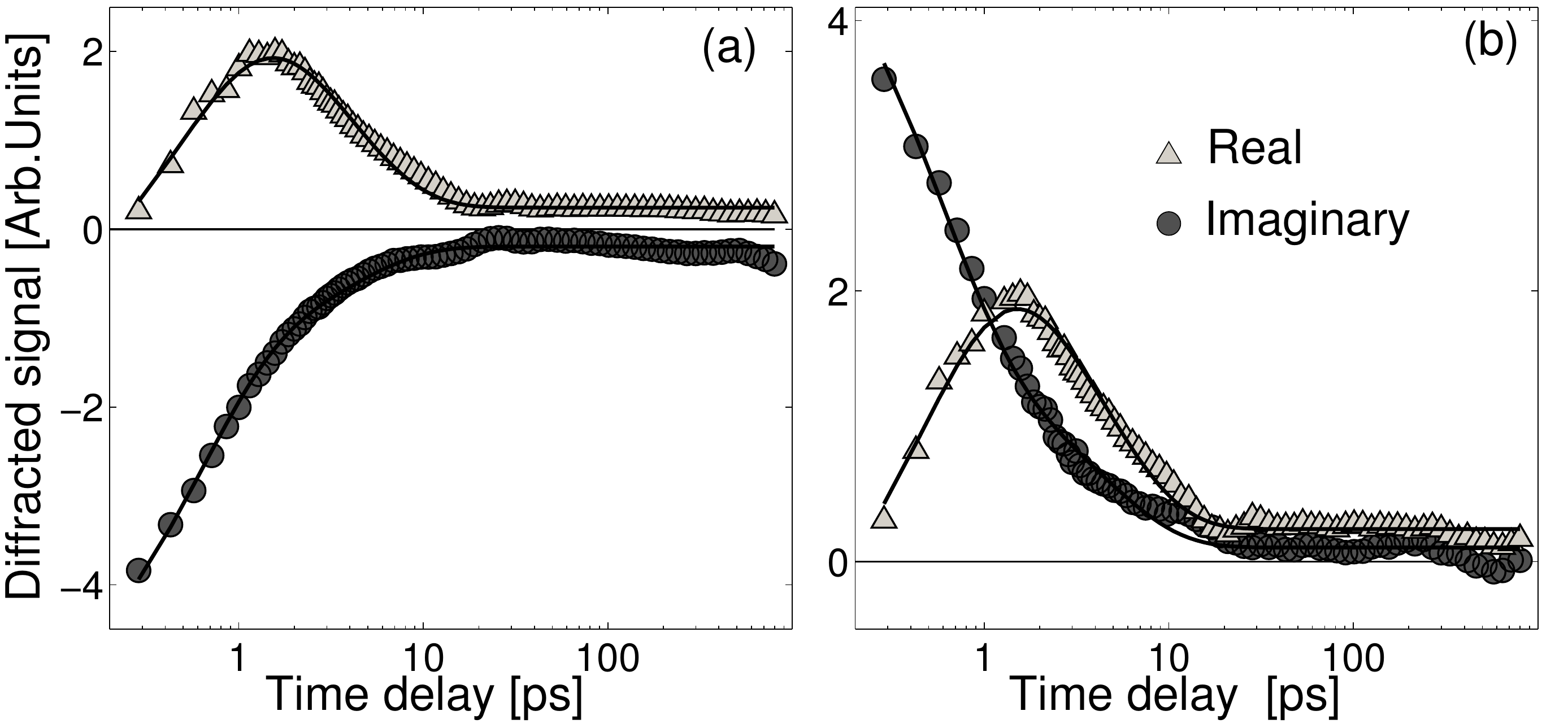}
\caption{Typical transient-grating data (semilog time). All three components of the signal are clearly visible. Real part, triangles; imaginary part, circles; lines are fits to Eq. \ref{twoexp}. $T=295$ K, $q=3.14$ $\mu$m$^{-1}$, $f=7.8$ $\mu$J/cm$^2$. \textbf{(a), (b)} are $m=+1$, $-1$ diffracted orders, respectively.}
\label{transients}
\end{figure}

Examples of the data obtained appear in Fig. \ref{transients}. All of our data fit well to the form:  
\begin{equation}\Delta r(t)=Ae^{i\theta_A}e^{-t/\tau_A}+Be^{i\theta_B}e^{-t/\tau_B}+Ce^{i\theta_C}.
\label{twoexp}\end{equation}
The data's three most salient features are each evident. First, the signal returns to equilibrium through two distinct decay processes, the first with $\tau_A=500\pm35$ fs and the second with $\tau_B= 3.1\pm 0.1 $ ps.\protect{\cite{errorbars}} Second, the two decay processes differ distinctly in complex phase. Finally, as shown in Fig. \ref{taus}, the decays are insensitive to both $q$ and $f$. Of these observations, the complex phase will play the key role in our identification, below, of the causes of the two decay processes. 

\subsection{Constancy of fit parameters}

\begin{figure}
\includegraphics[width=3.375 in, height=1.3 in]{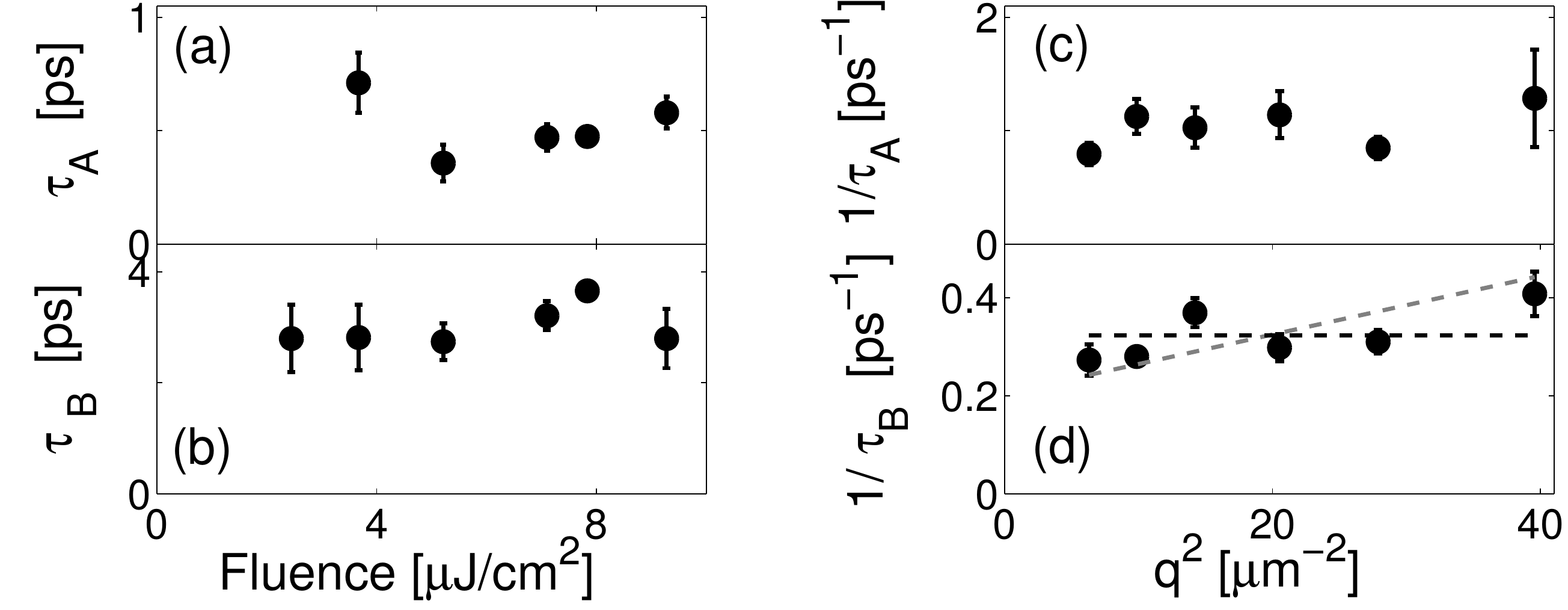}
\caption{\textbf{(a)} and \textbf{(b)}: $\tau_A$ and $\tau_B$ are roughly constant \textit{vs}. pump fluence. \textbf{(c)} and \textbf{(d)}: $1/\tau_A$ and $1/\tau_B$ are roughly constant \textit{vs}. $q^2$. $\tau_B$ is consistent with diffusion coefficients from $D=0$ (horizontal line) to $D=60$ cm$^2$/s (sloped line).}
\label{taus}
\end{figure}

In fact, the transient reflectance is even less sensitive to experimental conditions than Fig. \ref{taus} indicates. We varied the conditions---sample, $T$, $f$, $m$, and $q$---to measure 32 distinct $\Delta r(t)$ curves; we saw little variation in any of the fitting parameters of Eq. \ref{twoexp}. The relative size of the two decay processes is constant, $A/B=1.9\pm0.2$. The constant term increases from $C/B=0.05$ at 80 K to $0.10$ at 295 K, but always remains small. We attribute the $C$ term to lattice heating, for which we present qualitative evidence in the Supplemental Material.\protect{\cite{suppl}}

Transient-grating experiments are often used to measure the diffusivity $D$ of photoexcited species. In the presence of diffusion, the diffracted signal $d(t)$ decays faster than $\Delta r(t)$ because carriers diffuse from the grating's peaks to its troughs. This effect is stronger at higher $q$, because the peak-to-trough distance is shorter. However, Fig. \ref{taus} (d) shows that $\tau_B$ is independent of $q$, consistent with $D=0$. We caution against assigning too much weight to this negative result. The sloped line in Fig. \ref{taus} (d) shows that our data exclude only $D>60$ cm$^2$/s---a distinctly high upper bound. So the carriers likely do diffuse, but relax so quickly that they do not diffuse through an appreciable fraction of the grating's wavelength.\protect{\cite{ambipolar}} The situation for $\tau_A$ is similar: Fig. \ref{taus} (c). 

\subsection{Determination of absolute phase angles}
\begin{figure}
\includegraphics[width=2.86 in, height=1.5 in]{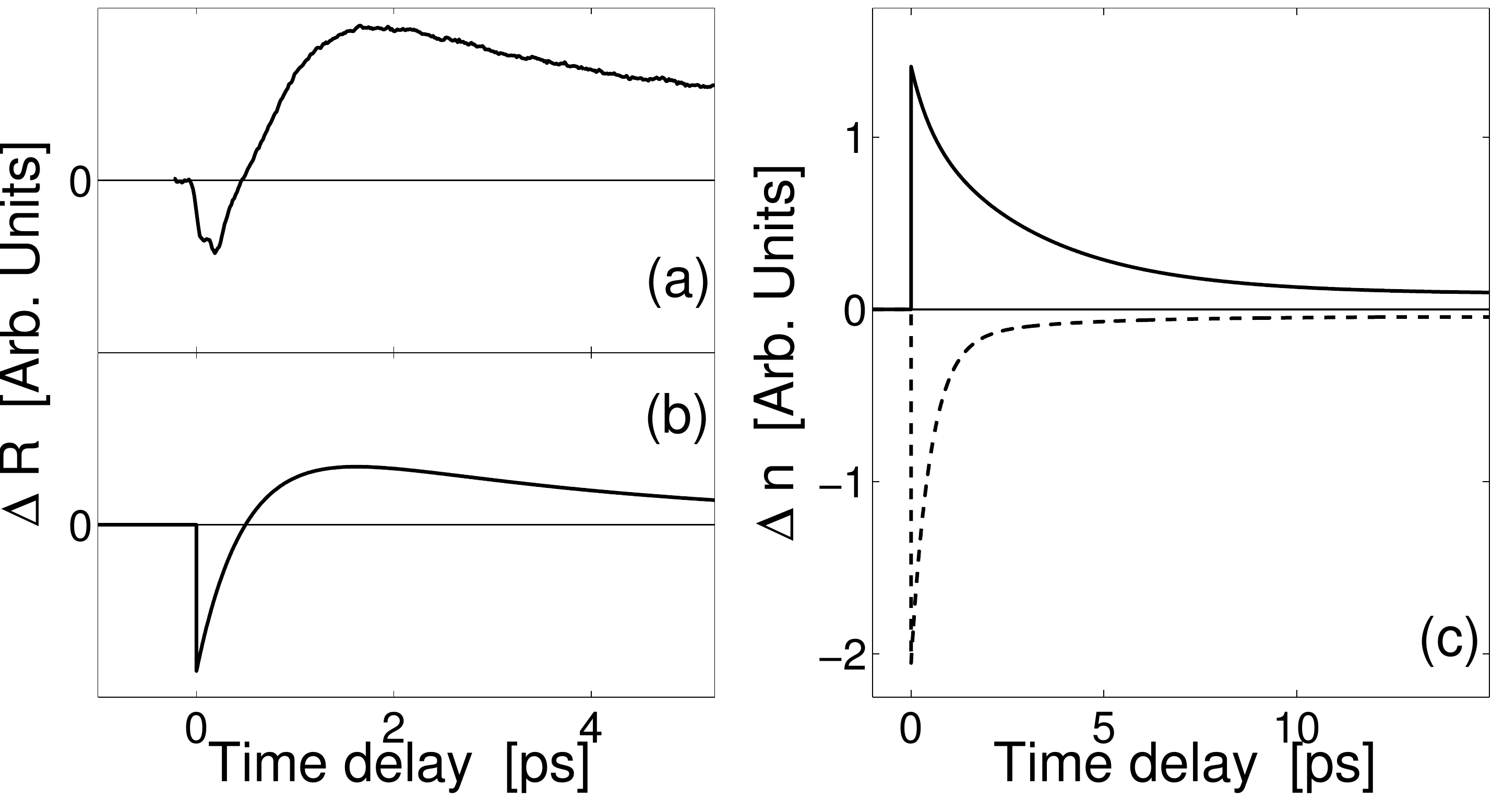}
\caption{\textbf{(a)}: Typical transient change in reflectivity, $\Delta R(t)$, measured. \textbf{(b)}: $\Delta R(t)$, calculated from our mean fit parameters. The sign of each component is chosen to match the shape of the measured curve.  \textbf{(c)}: Transient change, $\Delta n$, in index of refraction calculated from our mean fit parameters. Imaginary part, dashed, accounts for most of the fast decay. Real part, solid, accounts for most of the slow decay and the constant term.}
\label{phasefig}
\end{figure}

Our typical measurement, of $m=+1$, is not sensitive to the multiplication of Eq. \ref{twoexp} by an overall phase. However, by additionally measuring $m=-1$, it is possible to determine the absolute phase\protect{\cite{Gedik2004}} of $\Delta r$ . We have done several such measurements on each sample; one appears in Fig. \ref{transients} (b). We can then calculate\protect{\cite{suppl}}
\begin{equation}
\phi_{\Delta r}^A=\frac{\theta^{(-1)}_A-\theta^{(+1)}_A-\pi}{2}+\phi_r,
\label{halfangle}\end{equation}
and similarly for the signal's $B$ and $C$ components. Though the half-angle in Eq. \ref{halfangle} can take two values differing by $180\degree$, this ambiguity is easily resolved. The photoinduced change in reflectivity is $\Delta R=2|r||\Delta r(t)|\cos(\phi_r-\phi_{\Delta r})$; we measure $\Delta R(t)$ and choose the angles $\phi_{\Delta r}$ to reproduce its sign, shown in Figs. \ref{phasefig} (a) and \ref{phasefig} (b). 

We now use these angles to determine the photoinduced change in $n$. The reflectance changes after photoexcitation by $\Delta r(t) = [-2/(1+n)^2]\Delta n(t)$. For cadmium arsenide, the bracketed factor has argument $143\degree$, so $\phi_{\Delta n}=\phi_{\Delta r}-143\degree$. We obtain, finally, $\phi_{\Delta n}^A = -80\degree$, $\phi_{\Delta n}^B =  -8\degree$, and $\phi_{\Delta n}^C =  -25\degree$. 

This result is surprisingly simple. The signal's faster component results from a negative $\Delta n_i$---a reduction in absorption---and the slower from a positive $\Delta n_r$---a decrease in the light's phase velocity. The calculated $\Delta n(t)$ appears in Fig. \ref{phasefig} (c). For Cd$_3$As$_2$, both the real and imaginary parts of $\Delta n$ appear in $\Delta R$, and they may be distinguished by the time-scales of their decays. 

\section{Discussion}
The key questions in interpreting these two decay processes are what has been excited, and by what means it relaxes. Our excitation energy  $\varepsilon_p=1.5$ eV is well beyond the region of cadmium arsenide's Dirac-like dispersion, and, though optical transitions near 1.5 eV are believed to occur at the $\Gamma$ point,\protect{\cite{Karnicka1982}} transitions are allowed between electrons and holes of several different bands. Cadmium arsenide's large unit cell hosts over 200 phonon branches; infrared and Raman measurements detect a few dozen,\protect{\cite{Gelten1982, Jandl1984, Schleijpen1984, Houde1986}} with energies from 3.2 meV to 49 meV. (The deficit of detected branches is attributed to a weak polarizability.\protect{\cite{Houde1986}}) Considering the abundance of excited states and relaxation pathways available, we cannot hope to identify precise processes of excitation or relaxation. Nonetheless, the optical signal's phase constrains our interpretation significantly. 

Photoexcitation changes a sample's reflectance by changing its frequency-dependent absorption coefficient. Leaving aside the possibility of changes to the band structure, it does so either by occupying excited states or by changing the free carriers' absorption. Our experiment's probe photons have the same energy $\varepsilon_p$ as those of the pump. Therefore excited electrons fill phase-space effectively, reducing absorption at $\varepsilon_p$, and causing the negative $\Delta n_i$ observed in our fast decay process. 

This picture remains valid even as electrons scatter away from their initial excited energy $\varepsilon_e$.\protect{\cite{epsilon}} Carrier-carrier scattering gradually creates a thermal distribution of electrons at elevated temperature. If this process is fast compared to the carriers' energy loss, their mean energy remains nearly $\varepsilon_e$, and they occupy  states both below and above $\varepsilon_e$. Such a distribution results in $\Delta n_i<0$, just as does the conceptually simpler case of phase-space filling exactly at $\varepsilon_e$.

Our signal's slower component has $\Delta n_r>0$, which, according to the Kramers-Kronig relation, must result either from increased absorption at $\varepsilon>\varepsilon_p$ or from decreased absorption at $\varepsilon<\varepsilon_p$. We can eliminate the former as the cause of our signal. If absorption increases at all, it should do so at low frequency due to enhanced free-carrier (intraband) conductivity; this would cause a negative $\Delta n_r$ that we do not observe. On the other hand, there is a straightforward mechanism for decreased absorption at $\varepsilon<\varepsilon_p$: as electrons and holes lose their excess energy, they fill phase space at progressively lower energies. Kramers-Kronig analysis using a simplified density of states suggests that, by the time $\Delta n(\varepsilon_p)$ becomes mostly real, the carriers' mean energy should be $\varepsilon_e/2$ or less; our data show that cooling of this magnitude occurs within 500 fs. We attribute this cooling to phonons rather than to carrier-carrier scattering, since there are too few cool, background electrons compared to the hot, photoexcited ones (an order of magnitude fewer for Sample 1 and at our highest fluence).

The subsequent dynamics of $\Delta n_r$ indicates that once carriers reach low energy, their relaxation slows to give $\tau_B=3.1$ ps. Possibly cooling slows when the carriers' excess energy falls below that of the lowest optical phonon, as occurs in graphene\protect{\cite{Bistritzer2009, Strait2011}} and as recently preicted for Weyl and 3D Dirac semimetals.\protect{\cite{Lundgren2015}} However, for Cd$_3$As$_2$ this energy is just 15 meV.\protect{\cite{Houde1986}} Other possible relaxation processes include electron-impurity scattering or electron-electron scattering with plasmon emission. However, we suggest that after the initial 500-fs cooling the carriers and optical phonons have equilibrated; further cooling requires the slower emission of acoustic phonons. This picture fits the measured time-scale: electron-lattice cooling in bismuth, a semimetal, occurs in 5 ps.\protect{\cite{Sheu2013}}

We may gain insight into the two decay processes we observe in cadmium arsenide by considering another Dirac semimetal, graphene. Photoexcitation of graphene initially produces electrons and holes with separate chemical potentials.\protect{\cite{Gilbertson2012}} Within the pulse duration, these carriers partially equilibrate with optical phonons;\protect{\cite{LuiPRL2010}} they then quickly occupy the Dirac cone and enhance the intraband conductivity,\protect{\cite{Dani2012}} and recombine in less than a picosecond.\protect{\cite{Gilbertson2012}} The chemical potential reverts to its original level, but because carriers are still hot they continue to occupy high-energy states, filling phase-space and reducing optical absorption.\protect{\cite{Sun2008}} These hot carriers finally relax \textit{via} optical, then acoustic, phonons.\protect{\cite{Bistritzer2009, Strait2011}} 

Our measurements indicate that some of the same processes occur in cadmium arsenide, but possibly not all. We do not know whether carriers relax into the Dirac cone, but the weakness of cadmium arsenide's photoluminescence\protect{\cite{ONeil1990}} suggests that many do. We also cannot conclude that, as in graphene, photoexcitation produces electrons and holes with separate chemical potentials; time-resolved photoemission and THz could more directly detect changes in carrier population and conductivity. 

In conclusion, we have shown that after photoexcitation cadmium arsenide relaxes in two distinct stages, irrespective of sample, fluence, and temperature. First, carriers fill phase-space at the pump energy, but relax within 500 fs to lower energy. These low-energy carriers relax further with a time-scale of 3.1 ps; the lattice finally reaches high temperature. This result may guide further ultrafast measurements on Cd$_3$As$_2$ and other Dirac and Weyl semimetals.

\section{Acknowledement}

This work was supported by the National Science Foundation Grant No. DMR-1105553.
\begin{appendix}

\renewcommand\thefigure{A\arabic{figure}} 
\renewcommand\theequation{A\arabic{equation}} 

\section{Sample properties}

\begin{figure}
\includegraphics[width=3 in, height= 2in]{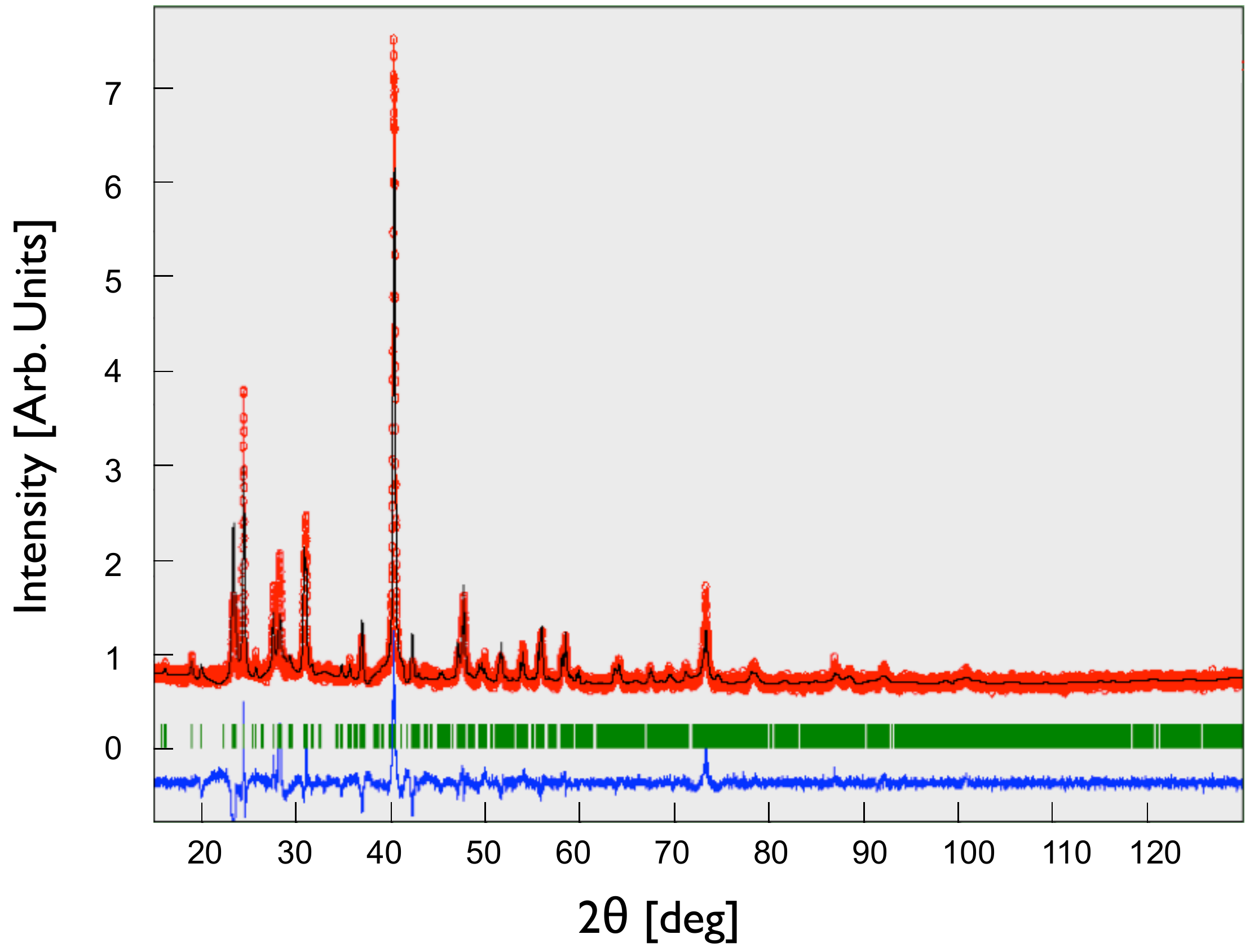}
\caption{X-ray diffraction pattern of powdered samples of type 1. \textbf{Red circles}: experimental data; \textbf{Black line}: calculated fit; \textbf{Blue line}: difference between fit and data; \textbf{Green bar}: Bragg positions.}
\label{xray}
\end{figure}

Here we describe further our two samples of Cd$_3$As$_2$. Fig. \ref{xray} shows the X-ray powder diffraction pattern from samples of the same vintage as Sample 1. The data were fit using Rietveld refinement, giving lattice parameters $a=b=12.6539$ {\AA} and $c=25.4586$ {\AA} with space group symmetry $I4_1/acd$, consistent with other recent experiments.\protect{\cite{Ali}} No peak corresponding to an impurity phase was detected. 

\begin{figure}
\includegraphics[width=3.375 in, height=1.2 in]{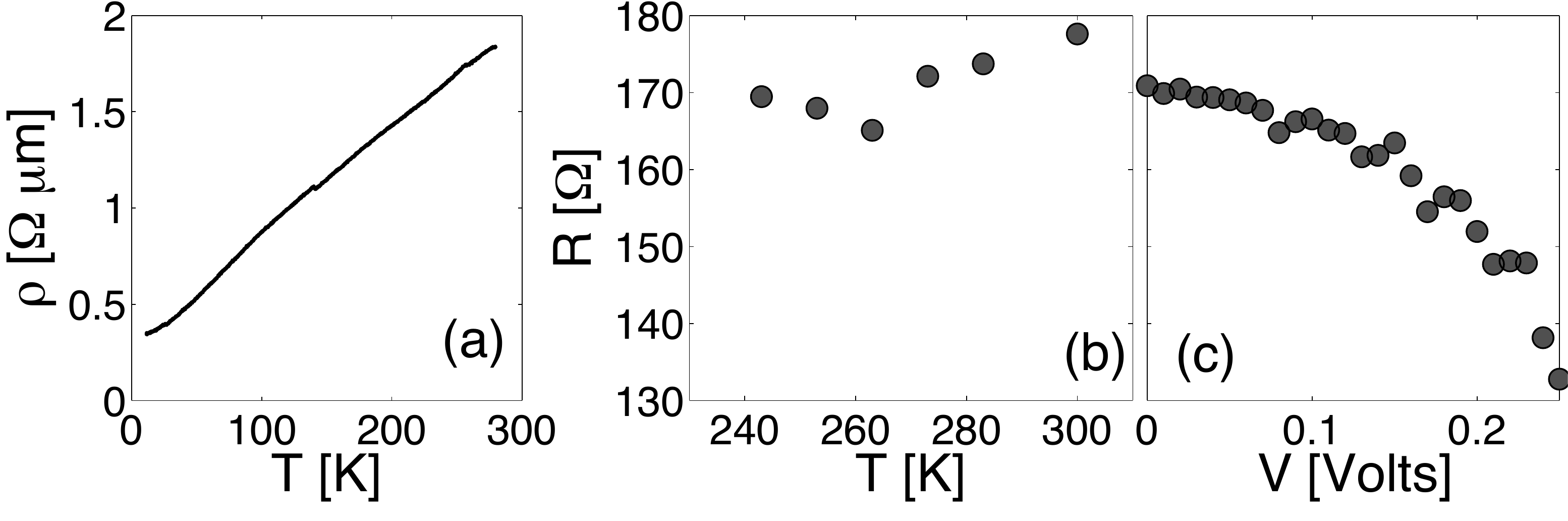}
\caption{\textbf{(a)}: Resistivity $\rho(T)$ for a sample of type 1. For a sample of type 2, we show \textbf{(b)} the zero-bias resistance $R_0(T)$ and \textbf{(c)} the differential resistance $R$ \textit{vs}. voltage $V$, averaged over the temperatures measured.}
\label{SampleDiagnostics}
\end{figure}

Fig. \ref{SampleDiagnostics} (a) shows the resistivity of a sample of the same vintage. The resistivity is metallic, and at low temperature is nearly as small as that measured in samples exhibiting confirmed Dirac-semimetal behavior.\protect{\cite{He2014}} 

For a sample of the same vintage as Sample 2, the current-voltage characteristics were measured by placing it across a gap formed by two indium electrodes. The DC photoconductivity at 300 K was negative, consistent with the optical heating of free carriers. The differential resistance, Fig. \ref{SampleDiagnostics} (b) and (c), is not unambiguously metallic: it depends non-monotonically on temperature and is bias-dependent. Nonetheless, in our experiment Samples 1 and 2 behaved identically.  

\section{Description of the transient grating}

Though transient-grating spectroscopy has a long history,\protect{\cite{Eichler1986}} the method has advanced considerably in recent decades.\protect{\cite{Vohringer1995, Chang1995, Goodno1998, Maznev1998, Maznev1998b, Gedik2004}} Here we provide additional details of the method and analysis used in this work.

\subsection{Mathematical description of heterodyne detection}

Using the notation of Eq. \ref{notation}, the reflected and diffracted fields are, respectively,
\begin{eqnarray}
E_r & = & |r|e^{i\phi_{r}}E_0+|\Delta r(t)|e^{i\phi_{\Delta r}}E_0,\nonumber\\
E_d & = & |d(t)|e^{i(\phi_{\Delta r}+m\phi_x)}E_0.
\end{eqnarray}

It is instructive to compare heterodyne detection to the traditional, homodyne-detected transient-grating signal, in which one simply measures the diffracted beam, $|E_d|^2$. The homodyne signal is second-order in $d(t)$. Unfortunately, the photoinduced changes in a sample's optical response---$d(t)$ and $\Delta r(t)$---are typically quite small. One advantage of heterodyne detection is a large increase in signal, for it has several terms of first or zeroth order:
\begin{eqnarray}
|E_r+E_d|^2=&&|E_0|^2(|r|^2+|\Delta r(t)|^2+|d(t)|^2\nonumber\\&+&2|r||\Delta r(t)|\cos(\phi_r-\phi_{\Delta r})\nonumber\\&+&2|r||d(t)|\cos(\phi_r-\phi_{\Delta r}-m\phi_x)\nonumber\\&+&2|d||\Delta r(t)|\cos(\phi_{\Delta r}+m\phi_x-\phi_{\Delta r})).\nonumber\\
\end{eqnarray}
Indeed, the second-order terms are negligible, giving
\begin{eqnarray}
|E_r+E_d|^2=&&|E_0|^2(|r|^2\nonumber\\&+&2|r||\Delta r(t)|\cos(\phi_r-\phi_{\Delta r})\nonumber\\&+&2|r||d(t)|\cos(\phi_r-\phi_{\Delta r}-m\phi_x).
\end{eqnarray}

The term of interest is the one proportional to $d(t)$; it is also the only one depending on $\phi_x$. To isolate this term, we use the coverslip discussed below to modulate $\phi_x$ at 95 Hz,\protect{\cite{Weber2005}} and filter our signal through a lock-in amplifier. This procedure acts as a derivative $d/d\phi_x$, giving a signal proportional to
\begin{equation}
-m|r||d(t)|\sin(\phi_r-\phi_{\Delta r}-m\phi_x),
\end{equation}
equivalent to Eq. \ref{eq2}. 

\subsection{Modulation of $\phi_x$}

Key to the heterodyne detection of the transient grating is the ability to control and modulate $\phi_x$, by controlling the grating's spatial position. We do this by passing one of the incident pump beams obliquely through a thin, glass coverslip. Fine adjustments of the coverslip's angle change the beam's path-length, adding or subtracting phase relative to the other pump beam. The coverslip is mounted on both a torsional oscillator and a stepping rotation stage. The former allows us to modulate the coverslip's angle rapidly and sinusoidally, for lock-in detection; the latter allows us to change the angle in calibrated increments. Below, when we discuss measurement ``at'' a particular coverslip position, we mean the coverslip's \textit{central} position, about which it oscillates at 95 Hz.

To maintain the spatial and temporal overlap of the beams that converge on the sample, we introduce three similar coverslips into the paths of both probe beams and of the other pump. Their orientations are fixed, but are similar to that of the modulated coverslip. 

To obtain data such as that shown in Fig. \ref{transients} (a), we set the coverslip to a position corresponding to an arbitrary, unknown $\phi_x$, and measure using the $m=+1$ diffracted probe. We then shift to $\phi_x+\pi/2$ and measure again. We obtain
\begin{equation}
|d(t)|\sin(\phi_r-\phi_{\Delta r}-\phi_x)
\label{real}\end{equation}
and
\begin{equation}
-|d(t)|\cos(\phi_r-\phi_{\Delta r}-\phi_x),
\label{imaginary}\end{equation}
and define these, respectively, as the real and imaginary parts of our signal: Re$(d(t))$ and Im$(d(t))$. Fitting to the form:  
\begin{equation}d(t)=Ae^{i\theta_A}e^{-t/\tau_A}+Be^{i\theta_B}e^{-t/\tau_B}+Ce^{i\theta_C}
\label{twoexp}\end{equation}
defines the set of angles $\theta_A^{(+1)}$, $\theta_B^{(+1)}$, $\theta_C^{(+1)}$. The superscript indicates that $m=+1$. Hereafter we consider just one of the signal's $A$, $B$, and $C$ components; the same equations apply to each. 

Comparison of Eqs. \ref{real}, \ref{imaginary}, and \ref{twoexp} shows that 
\begin{equation}\theta^{(+1)}=(\phi_r-\phi_{\Delta r}-\phi_x)-\frac{\pi}{2},
\end{equation}
because Re$(d(t))\propto\cos\theta^{(+1)}$ and Im$(d(t))\propto\sin\theta^{(+1)}$.

\subsection{Determination of phase angles}

Up to this point, $\theta$ is arbitrary, because $\phi_x$ is arbitrary. We next describe how measurement of the $m=-1$ diffracted order allows us to eliminate $\phi_x$ and to determine $\phi_{\Delta r}$. 

Our measurements of the $m=-1$ diffracted order are done at the same coverslip positions as for the $+1$ order. These correspond to grating phases of $\phi_x$ and $\phi_x-\pi/2$. Our transient-grating signals are, respectively, 
\begin{equation}
|d(t)|\sin(\phi_r-\phi_{\Delta r}+\phi_x)
\label{real-1}\end{equation}
and
\begin{equation}
|d(t)|\cos(\phi_r-\phi_{\Delta r}+\phi_x),
\label{imaginary-1}\end{equation}
from which
\begin{equation}\theta^{(-1)}=\frac{\pi}{2}-(\phi_r-\phi_{\Delta r}+\phi_x).
\end{equation}

We can then calculate
\begin{equation}
\phi_{\Delta r}=\frac{\theta^{(-1)}-\theta^{(+1)}-\pi}{2}+\phi_r,
\label{halfangle}\end{equation}
equivalent to Eq. \ref{halfangle}. 

\subsection{$q=0$ limit}

Above, we state that $\Delta r(t)=d(t)$ because the diffusion is negligible. Here we clarify the reasoning, which may otherwise appear circular. $\Delta r(t)$ must equal the $q=0$ limit of $d(t)$. We observe that, within experimental error, $d(t)$ is the same at all $q$, including at some rather low $q$. Therefore our measured $d(t)$ does in fact equal $\Delta r(t)$. In other words, we never assume that the diffusion is negligible; we observe it.

\subsection{Diffusion into the bulk}
\begin{figure}
\includegraphics[width=2.75 in, height=1.6 in]{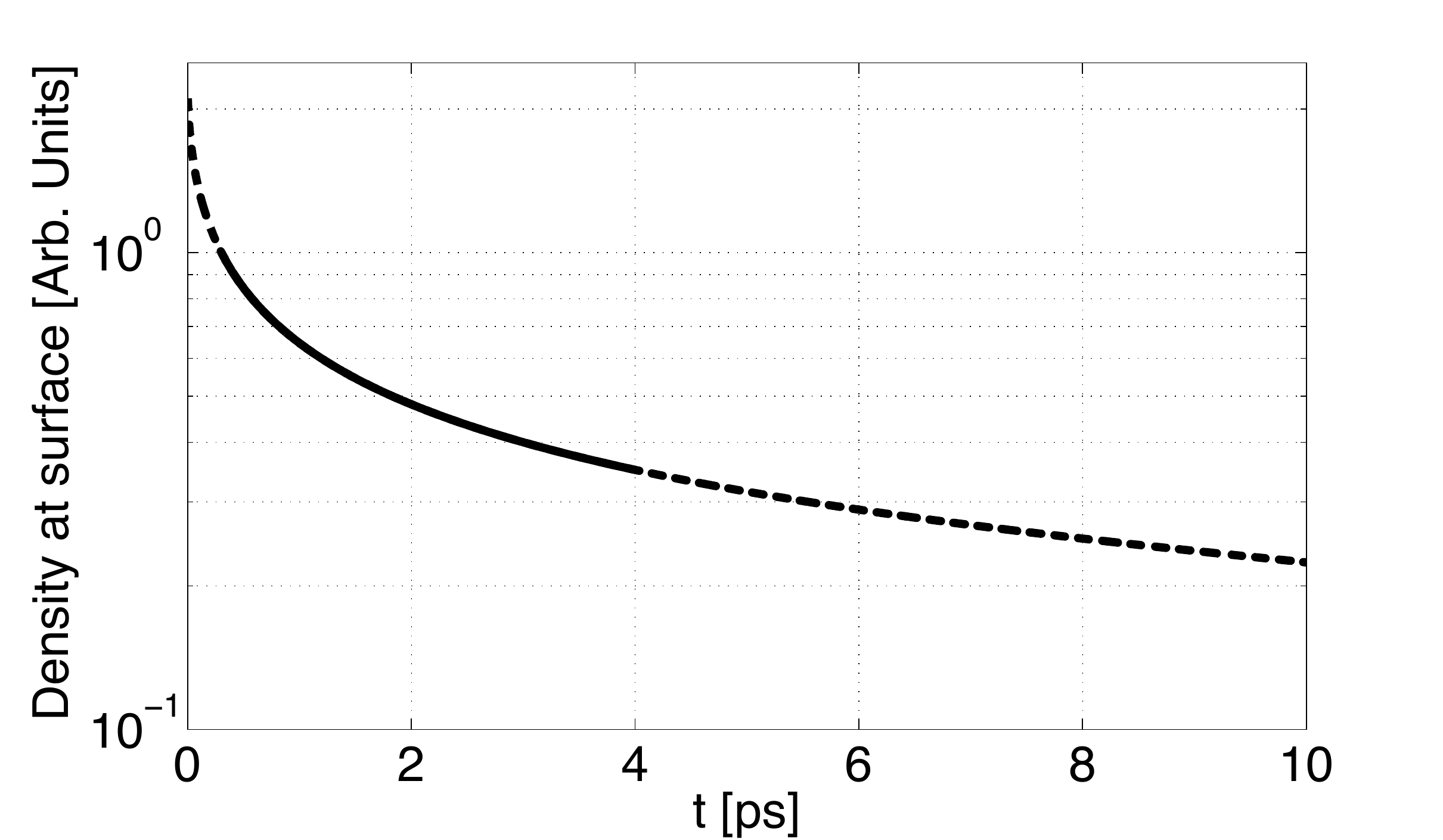}
\caption{Calculated density of photoexcited carriers at the sample's surface, for a model including diffusion into the bulk but excluding recombination. The solid line extends from 0.3 ps (at which our experimental data begin) to 4 ps, during which time our observed signal decreases by a factor of 7.1. During the same time, diffusion decreases the carrier density by a factor of just 2.9. (The calculation uses $L=45$ nm and $D=60$ cm$^2$/s. If $D$ is lower, the density will be even more nearly constant.)}
\label{inward}
\end{figure}

Because carriers are photoexcited within an absorption length $L$ of just 45 nm, they will diffuse away from the sample's surface. As their density at the surface drops, $\Delta n$ will decrease. Surprisingly, however, this effect has little influence on the time constants $\tau_A$ and $\tau_B$ measured in this experiment. The carriers' initial exponential distribution quickly evolves to be nearly Gaussian, and diffusion broadens a Gaussian's width only as $\sqrt{t}$. (See, for instance, Sheu \textit{et al.},\protect{\cite{Sheu2013}} particularly Eq. 3 and Fig. 5a.)

Fig. \ref{inward} makes this argument more quantitative. We used the diffusion equation with no relaxation term---\textit{i.e.} for conserved particle number---to model carriers' diffusion away from the sample's surface, and we plot the density at the surface as a function of time.\protect{\cite{t0}} The largest drop in density occurs at very early times, before any of the data shown in Fig. \ref{transients}. The drop at later times is not nearly enough to account for the experimentally observed decays. For this reason we conclude that our relaxation rates are little influenced by inward diffusion.

\section{Optical heating}

\subsection{Heating of electrons}

Above, we measure the photoinduced change of reflectance, $\Delta r(t)$, and argue that it is related to the cooling of optically heated electrons. Here we describe qualitative evidence for the electrons' high temperature.

When the laser was incident on the sample, we saw that the illuminated spot glowed with broadband visible light; it looked like incandescence, of a reddish hue. Some locations on the sample surface glowed more than others; however, we excluded surface contamination as a cause of the light emission by visual examination of the sample and by cleaning with acetone and methanol. Lui \textit{et al}.\protect{\cite{LuiPRL2010}} measured a similar effect in graphene, caused by thermal emission by electrons heated to several thousand Kelvin. These electrons were partially equilibrated with the optical phonons. After equilibration with all phonon modes, the lattice temperature was estimated to be around 700 K. The more complex band structure of Cd$_3$As$_2$ precludes the quantitative analysis of Lui \textit{et al}., but we expect that emission from our sample is caused by similarly heated electrons. 

\subsection{Heating of the lattice}

Above, we attribute our signal's small, constant component $C$ to lattice heating. It is unremarkable that energy deposited by the laser should eventually find its way to the lattice. However, given the modest optical power used in our experiment---tens of milliwats for the pump beams---this heating maifests itself in a surprising way that may serve to caution future experimenters.

We observed that the direction of the probe beam's specular reflection from the sample's surface could vary by about $2\degree$, depending on whether the more powerful pump beam was incident on the sample or blocked. This change was reproducible over dozens of cycles, and occurred with a time constant of several seconds. Reflection remained specular, but the orientation of the sample's surface evidently shifted. After many cycles, Sample 1's surface showed small cracks. 

We explain this observation as follows. The thermal conductivity of cadmium arsenide is low,\protect{\cite{Armitage1969}} of order 1 W/K-m, leading to large temperature gradients. The material suffers several structural phase transitions at elevated temperature,\protect{\cite{Zdanowicz1975}} the lowest at 503 K; combined with temperature gradients, these could create strains that move the sample's surface slightly.   

Note that a transient-grating signal cannot be measured when the reflected beam is shifting. We were able to obtain data because samples neither shifted nor glowed when exposed to atmosphere, perhaps due to convective cooling. For low-temperature measurement under vacuum, Sample 2 glowed and shifted only rarely, evidently depending on which part of the sample was illuminated. The sample was never measured while incandescing; nonetheless, it is likely that even when the sample was cooled to 80 K, the measured spot was much hotter. Both a second platelet-like sample and Sample 1 glowed and shifted more consistently and could not be measured under vacuum. These observations suggest the value of thin-film samples or of laser systems with lower average power. 
\end{appendix}


\begin{thebibliography}{10}
\bibitem{errorbars}
Errors and error-bars are standard deviation of the mean.
\bibitem{ambipolar}
The lack of measurable diffusion is perhaps to be expected. Photoexcited electrons and holes must move together to preserve local charge neutrality, resulting in ambipolar diffusion. In an $n$-type sample, the ambipolar diffusivity nearly equals the holes' diffusivity, which is likely much less than the electrons'.
\bibitem{epsilon}
Note that electronic energy of $\varepsilon_e$ corresponds to optical transitions at $\varepsilon_p$.
\bibitem{t0} The characteristic time-scale of the diffusive process is $t_0=L^2/D$. In this experiment $t_0\geq0.3$ ps. Indeed, this is the \textit{only} time-scale in the problem, so that the $t$-axis in Fig. \ref{inward} may we re-scaled by the factor $t_0/0.3$ to adapt it to other experiments.
\end{thebibliography}
\end{document}